\documentclass[]{aastex631} %linenumbers

\usepackage{CJK}
\usepackage{booktabs}
\usepackage{color}
\usepackage{tablefootnote}
\usepackage{graphicx}	% Including figure files
\usepackage{subfigure}
\usepackage{float}
\usepackage{amsmath}	% Advanced maths commands
\usepackage{amssymb}	% Extra maths symbols
\usepackage{ltxtable}
\usepackage[normalem]{ulem}
\usepackage{amsmath,amstext,color,tabu,mathtools}
\usepackage{gensymb}
\usepackage{natbib}
\usepackage{hyperref}
\usepackage{longtable}
\usepackage{mathrsfs}
\usepackage{comment}
\usepackage{footmisc}
\usepackage{subfigure}

\begin{document}
\begin{CJK*}{UTF8}{gbsn}

\title{A Novel Model for the MeV Emission Line in GRB~221009A}

%\correspondingauthor{Da-Ming Wei, Zi-Gao Dai}
%\email{daizg@ustc.edu.cn, dmwei@pmo.ac.cn}
%\email{dmwei@pmo.ac.cn}

\author[0000-0002-9775-2692]{Yu-Jia Wei (魏煜佳)}
\affiliation{Key Laboratory of Dark Matter and Space Astronomy, Purple Mountain Observatory, Chinese Academy of Sciences, Nanjing 210034, China; dmwei@pmo.ac.cn; yjwei@pmo.ac.cn}
\affiliation{School of Astronomy and Space Science, University of Science and Technology of China, Hefei, Anhui 230026, China}

\author[0000-0002-9037-8642]{Jia Ren (任佳)}
\footnote{Yu-Jia Wei and Jia Ren contributed equally to this work and should be considered co-first authors.}
\affiliation{School of Astronomy and Space Science, Nanjing University, Nanjing 210093, China; jia@smail.nju.edu.cn}

\author[0000-0002-8941-9603]{Hao-Ning He (贺昊宁)}
\affiliation{Key Laboratory of Dark Matter and Space Astronomy, Purple Mountain Observatory, Chinese Academy of Sciences, Nanjing 210034, China; dmwei@pmo.ac.cn; yjwei@pmo.ac.cn}
\affiliation{School of Astronomy and Space Science, University of Science and Technology of China, Hefei, Anhui 230026, China}

\author[0000-0001-6374-8313]{Yuan-Pei Yang (杨元培)}
\affiliation{South-Western Institute for Astronomy Research, Yunnan University, Kunming 650504, China}
\affiliation{Purple Mountain Observatory, Chinese Academy of Sciences, Nanjing 210023, China}

\author[0000-0003-4977-9724]{Da-Ming Wei (韦大明)}
\affiliation{Key Laboratory of Dark Matter and Space Astronomy, Purple Mountain Observatory, Chinese Academy of Sciences, Nanjing 210034, China; dmwei@pmo.ac.cn; yjwei@pmo.ac.cn}
\affiliation{School of Astronomy and Space Science, University of Science and Technology of China, Hefei, Anhui 230026, China}

\author[0000-0002-7835-8585]{Zi-Gao Dai (戴子高)}
\affiliation{Department of Astronomy, School of Physical Sciences, University of Science and Technology of China, Hefei 230026, China; daizg@ustc.edu.cn}

\author[0000-0003-2478-333X]{B. Theodore Zhang (张兵)}
\affiliation{Center for Gravitational Physics and Quantum Information, Yukawa Institute for Theoretical Physics, Kyoto University, Kyoto 606-8502, Japan}



\begin{abstract}
Gamma-ray bursts (GRBs) have long been considered potential sources of ultra-high-energy cosmic rays (UHECRs; with energy $\gtrsim 10^{18} {\rm~eV}$). In this work, we propose a novel model generating MeV emission lines in GRB, which can constrain the properties of heavy nuclei that potentially exist in GRB jets. Specifically, we find that relativistic hydrogen-like high-atomic-number ions originating from the $\beta$ decay of unstable nuclei and/or the recombination entrained in the GRB jet can generate narrow MeV emission lines through the de-excitation of excited-electrons. This model can successfully explain the MeV emission line observed in the most luminous GRB ever recorded, GRB~221009A, with suitable parameters including a Lorentz factor $\gamma \sim 820-1700$ and a total mass of heavy nuclei $M_{\rm tot} \sim 10^{23} - 10^{26}$~g. Especially, the emission line broadening can be reasonably attributed to both the expansion of the jet shell and the thermal motion of nuclei, naturally resulting in a narrow width ($\sigma_{\rm line} / E_{\rm line} \lesssim 0.2$) consistent with the observation. Furthermore, we predict that different GRBs can exhibit lines in different bands with various evolving behaviors, which might be confirmed with further observations. Finally, our model provides indirect evidence that GRBs may be one of the sources of UHECRs.
\end{abstract}

\keywords{(stars:) gamma-ray burst: general -- line: formation}


\section{Introduction} \label{sec:intro}
Ultra-high-energy cosmic rays (UHECRs) are the most energetic particles ever observed with the energy $\gtrsim 10^{18} {\rm~eV}$, 
and are thought to be of an extragalactic origin due to the inability of the magnetic field in the Galactic disk to constrain particles of such a high energy~\citep[e.g.,][for a review]{Hillas_1984ARA&A..22..425H,Anchordoqui_2019PhR...801....1A}. 
There are several acceleration mechanisms to explain the origin of UHECRs, including magnetospheric models, jet models, and diffusive shock acceleration models~\citep[e.g.,][]{Globus_2023EPJWC.28304001G}.
Especially, gamma-ray bursts (GRBs), related to newborn stellar-mass black holes (BHs) generating relativistic jets, are considered to be one of the potential origins for UHECRs~\citep[e.g.,][]{Waxman_1995PhRvL..75..386W,Milgrom_1995ApJ...449L..37M,Vietri_1995ApJ...453..883V,Zhang_2018pgrb.book.....Z,Murase_2019ARNPS..69..477M,Globus_2023EPJWC.28304001G}. 

The surrounding environment around GRB jets~\citep[e.g.,][]{Prochaska_2007ApJ...666..267P,Fynbo_2009ApJS..185..526F} 
and their Lorentz factors~\citep[e.g.,][]{Nava_2017MNRAS.465..811N,Ghirlanda_2018A&A...609A.112G,Lithwick_2001ApJ...555..540L} imply that GRBs must contain at least some protons or even heavier nuclei since the simulated spectra for GRBs with only photons and $e^\pm$ pairs are inconsistent with observations~\citep[e.g.,][for a review]{Zhang_2018pgrb.book.....Z}. 
Observations of iron lines in the X-ray afterglow of GRBs also suggest the existence of iron nuclei associated with GRBs~\citep[e.g.,][]{Lazzati_1999MNRAS.304L..31L,Ballantyne_2001ApJ...559L..83B,Kallman_2003ApJ...593..946K}.
In addition, some theories proposed that GRB jets can carry heavy nuclei~\citep{Horiuchi_2012ApJ...753...69H, Zhang_2018PhRvD..97h3010Z}, which can be accelerated to an ultra-high-energy (UHE).
Both high-luminosity long GRBs and low-luminosity GRBs have been considered as potential sources of
UHECRs~\citep[e.g.,][for a review]{Zhang_2018pgrb.book.....Z}. 
%\yj{However, the direct and airtight observation evidence is still lacking. }
%The GRB - UHECR proton connection from TeV photons in GRB 221009A may be questioned by the energy reconstruction in the LHAASO data.}
Some works have studied the spatial coincidence between UHECRs and GRBs~\citep{Milgrom_1995ApJ...449L..37M,Mirabal_2023JCAP...02..047M}.
Recently, the observations of multi-TeV gamma rays from the brightest GRB 221009A suggest that cosmic ray protons have been accelerated to ultrahigh energies in the jet without considering exotic physics~\citep{Das_2023A&A...670L..12D,Kalashev_2024arXiv240505402K}.

In this paper, we propose a model which could generate the MeV emission line in GRB. Based on the observation of MeV emission lines, we can constrain the properties of possible existing heavy nuclei in GRB jets. 
It is the first indication that a GRB jet can accelerate nuclei (in addition to protons) to very high energies.
We invoke the excitation and de-excitation model in \cite{Kusenko_2012PhLB..707..255K}, but use it for relativistic "cold" high-atomic-number (high-Z) ions to produce the narrow MeV emission lines in the GRB prompt emission spectra, thereby limiting the parameters of GRBs and associated heavy nuclei based on emission line observations.
Recently, \cite{Edvige_2023arXiv230316223E} found a highly significant ($> 6 \sigma$) narrow line with an energy around 10~MeV and a luminosity around $10^{50}\ \rm erg\ s^{-1}$ in the brightest GRB~221009A ever observed. 
Although they suggested that it could be explained by the annihilation line of relatively cold $e^{\pm}$ pairs~\citep[e.g.,][]{Peter_2004ApJ...613..448P,Ioka_2007ApJ...670L..77I,Murase_2008ApJ...676.1123M}, this explanation faces challenges due to the required small Lorentz factor and the broad line width. However, our model can well explain the MeV emission line observed in the prompt stage of GRB~221009A. 
Our paper is structured as follows: In Sec.~\ref{sect:model}, we illustrate the model in detail. In Sec.~\ref{sect:calculation}, we show the calculations for nuclei in GRB jets based on the observed emission line. In Sec.~\ref{sect:221009a}, we use this model to explain the MeV emission line in GRB~221009A. In Sec.~\ref{sect:sum}, we summarize our results and make some predictions.

\section{Model for MeV Emission Lines} 
\label{sect:model}

\begin{figure}[hbtp]
    \centering
    \includegraphics[width=0.49\textwidth]{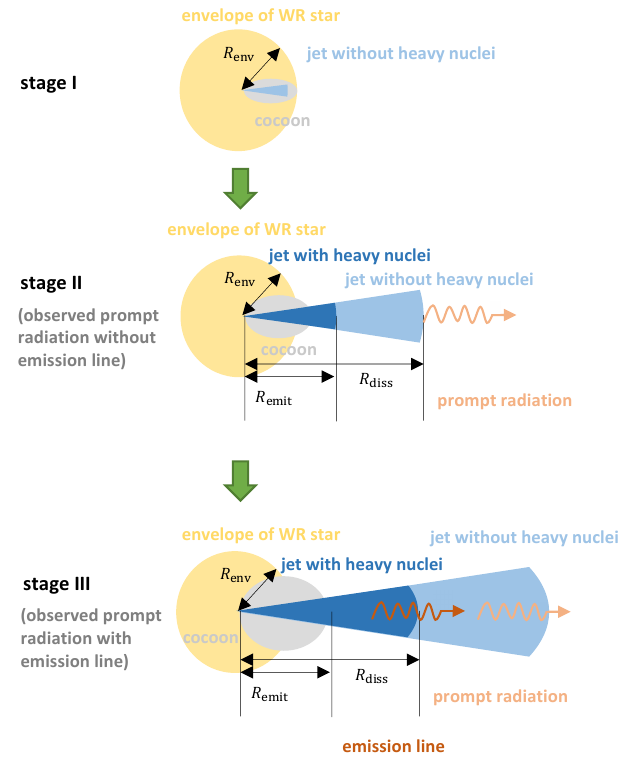}
    \label{fig:model2}
    
    \caption{
    A schematic picture for the generation of the MeV emission line. 
    The mass-dominated jet head (the jet without heavy nuclei) could not carry heavy nuclei due to its high temperature, which corresponds to effective photodisintegration and spallation. Once the jet becomes Poynting-dominated, it can entrain heavy nuclei.
    After the jet with heavy nuclei breaks through the photosphere and hydrogen-like heavy ions are generated by $\beta$ decay and/or recombination, the emission line can be generated at $R_{\rm emit}$. Around this point, the jet head reaches the dissipation radius $R_{\rm diss}$, generating the prompt emission. When the jet with heavy nuclei arrives at $R_{\rm diss}$, it cannot generate the emission line due to band broadening and/or spallation, leading to the disappearance of the emission line in the end. } 
    \label{fig:model}
\end{figure}

It is difficult for high-Z nuclei to survive in the initial fireball, as its high temperature can destroy the nuclei~\citep{Dermer_2007arXiv0711.2804D}. After expanding over some distance, they can be carried in the relativistic jet from the stellar core and/or the accretion disk wind of the hyperaccretion BH~\citep[e.g.,][]{Wang_2008ApJ...677..432W,Horiuchi_2012ApJ...753...69H, Zhang_2018PhRvD..97h3010Z}.
Particularly, although heavy nuclei cannot exist in a matter-dominated jet within $\sim 10^9$~cm for high-luminosity GRBs, they can survive in a Poynting-flux-dominated jet, as its lower temperature leads to more ineffective photodisintegration and spallation~\citep{Horiuchi_2012ApJ...753...69H,Zhang_2018PhRvD..97h3010Z}.
However, the bound electrons around these nuclei might be ionized since the typical temperature $\sim (1-10) ~\rm keV$ of the photonsphere~\citep[e.g.,][]{Peng_2014ApJ...795..155P} is higher than the ionization energy of electrons with energy states of $n > 1$ (e.g., $\lesssim 2$~keV for iron~\citep[e.g.,][]{NIST_ASD}), and is potentially sufficient even for $n = 1$ (e.g., $\sim 8$~keV~\citep[e.g.,][]{NIST_ASD} for iron). For simplicity, we consider these nuclei to be hydrogen-like ions.

According to the observations of GRB~160625B, they found a transition from a fireball to a Poynting-flux-dominated outflow. If this transition also exists in the process that we consider, it will cause a part of the jet without heavy nuclei (corresponding to the fireball stage), as shown in Fig.~\ref{fig:model}. 
Also, the dissipation process generating prompt emission at $R_{\rm diss}$ will destroy these ions by spallation and/or photodisintegrating process~\citep{Wang_2008ApJ...677..432W}. Besides, the survived-ions may be shock-accelerated to high random Lorentz factors thus no longer being 'cold'. 
Therefore, the hydrogen-like high-Z ions in the jet except for the jet head only can exist at a coasting phase from a characteristic radius $R_{\rm ion} \simeq c \gamma \tau_{\beta, 1/2}$ to $R_{\rm diss}$, where $\tau_{\beta, 1/2}$ is the $\beta$ decay half-life timescale of unstable nuclei at the comoving frame, and $\gamma$ is the Lorentz factor of nuclei (which is the same as the Lorentz factor $\Gamma$ of the coasting GRB jet). Note that this puts a requirement in our model on the unstable nuclei, that is $\tau_{\beta, 1/2} < R_{\rm diss} / (c \gamma) \sim (1 - 10^3) ~{\rm s}$ considering $\gamma = 10^{2} - 10^{3}$ and $R_{\rm diss} = (10^{13} - 10^{15})$~cm. 

We consider that the narrow MeV emission lines can be generated by the excitation and de-excitation of the hydrogen-like high-Z 'cold' ions in a relativistic jet with a Lorentz factor of $\gamma$. We assume the seed photons that excite the bound electrons are isotropic blackbody photons in the circumburst environment with a temperature of $T$. 
The seed photons with an energy of $\epsilon_\gamma \gtrsim \epsilon_z / \gamma$, considering the different direction between photons and ions (where $\epsilon_z = m_e \alpha^2 c^2 Z^2 / 2$ is the Rydberg energy), will excite the ground electrons to one of the higher bound states. The de-excitation of these excited electrons occurs practically instantaneously~\citep{Kusenko_2012PhLB..707..255K}. The seed photons also can ionize the bound electron in a timescale of $\sim 10^3$~s with typical parameters, which is much longer than the excitation timescale and the de-excitation timescale. 
The cross-section of recombination is $\propto n^{-3}$ (see Eq.~\ref{sigma_rb}), which means the recombination to the bound state ($n = 1$) is the most important. For typical parameters, the recombination timescale $\tau_r \sim 10^{-2}$~s is much smaller than the ionization timescale $\tau_i$. 
Therefore, the repeated cycles of excitation and radiation can generate the emission lines at an energy of $E_{\rm line} \simeq \gamma \epsilon_z$ within the propagation distance of $d$ in the lab frame.
Note that to satisfy the requirement that the iron can survive from both photodisintegration and nuclear spallation, the thermal motion temperature $T_t$ of the high-Z ion can not be higher than $\sim 10$~MeV~\citep{Wang_2008ApJ...677..432W}. We have $k_B T_t / (m_p c^2) \ll 1$ for the high-Z ions, which indicates that if the high-Z ion follows the Maxwell distribution, its velocity profile is narrow and naturally satisfies the condition of a narrow emission line. 
For the region generating emission lines, in addition to the survival region of the hydrogen-like high-Z ions in the jet, we need to consider the photosphere radius $R_{\rm ph}$ of GRBs (at which the optical depth is the order of one) since the photons within the photosphere radius are difficult to escape and may get heated into the blackbody radiation. Therefore, the emission line can only be generated at a radius from $R_{\rm emit}$ to $R_{\rm diss}$, where $R_{\rm emit} = \max \left(R_{\rm ion}, R_{\rm ph}\right)$.

The model we proposed could generate narrow MeV emission lines in the GRB prompt emission spectra, by which we can approximately constrain 
the total mass of the jet-carried heavy unstable nuclei ($M_{\rm tot}$) and the coasting phase Lorentz factor ($\gamma$) based on the emission line observations, as shown in Fig.~\ref{fig:model2}.
The jet head without high-Z ions arrives at $R_{\rm diss}$, leading to the prompt emission. Around this moment, the following jet with high-Z ions arrives at $R_{\rm emit}$ generating the emission line. 
The emission line photons will be delayed with a time $\Delta t_{\rm gap} \sim (R_{\rm diss} - R_{\rm emit}) (1 + z) / c$ of the initial prompt emission in the observer frame.
The jet with high-Z ions can generate emission lines from $R_{\rm emit}$ to $R_{\rm diss}$, but the shell that emits most photons will dominate observations. 
When all the jets with high-Z ions arrive at $R_{\rm diss}$ or the flux of emission lines is lower than the prompt emission, the emission lines will disappear.

\section{Properties of Nuclei in GRB Jets Derived from Observed Emission Lines}
\label{sect:calculation}

The energy of emitting photons at the observer frame is 
\begin{equation}
\label{eq:E_line}
    E_{\rm line} = \gamma \Delta_2/(1 + z),
\end{equation}
where $\Delta_2 = 3/4 \epsilon_z$ is the excitation energy from $n = 2$ to $n = 1$, and $z$ is the redshift of the GRB. Note that here we use the photons generated from the transition of electrons from $n = 2$ to $n = 1$ due to their maximum probability. We can then get the Lorentz factor $\gamma$ of the jet in the coasting phase from the observed energy $E_{\rm line}$ of the emission line.
The observed emission line width $\sigma_{\rm line}$ can be explained by the spreading of emitting shells during the coasting stage of the jet,
\begin{equation}
\begin{aligned}
&\gamma_{\rm rel}=1+\frac{\sigma_{\rm line}}{E_{\rm line}}, \\
&\beta_{\rm rel}=\sqrt{1-\gamma^{-2}_{\rm rel}},
\end{aligned}
\end{equation}
where $\beta_{\rm rel}$ is the expanding velocity of a spreading shell in the shell comoving frame. Theoretically, one may expect it to be consistent with the sound speed of a relativistic gas, that is $\beta_{\rm rel} \lesssim 1/\sqrt{3}$, naturally resulting in a narrow width ($\sigma_{\rm line} / E_{\rm line} \lesssim 0.2$).

Following the calculations in \cite{Kusenko_2012PhLB..707..255K}, considering the interaction with isotropic blackbody photons, the total excitation rate $\Gamma_e$ for an electron of the high-Z ion transitioned from the ground ($1S$) in the lab frame is
\begin{equation}
\Gamma_{e} = \frac{64 \alpha \pi^3 k_{\mathrm{B}}^3}{3 c^2 h^3} a_Z^2 T^3 X\left(\frac{\gamma}{\gamma_Z}\right)
\label{eq:Gamma_e}
\end{equation}
where
\begin{equation}
\begin{aligned}
& X(r) = 256 \sum_{n=2}^{\infty} \frac{n^7}{\left(n^2-1\right)^5}\left(\frac{n-1}{n+1}\right)^{2 n} \chi\left[\left(1-n^{-2}\right) / r\right], \\
& r = \gamma / \gamma_z, \\
& \chi(z) = -z^2 \ln \left(1-e^{-z}\right), \ z = \left(1-n^{-2}\right) / r.
\end{aligned}
\end{equation}
$a_Z = a_B / Z$ is the dipole matrix elements for a simple hydrogen-like ion, $Z$ is the atomic number, $a_B = h / (2 \pi m_e c \alpha)$ is the Bohr radius, $\gamma_z = \epsilon_z / (2 k_B T)$ is the characteristic value of the gamma factor for photons, and $T$ is the temperature for blackbody photons.
The typical timescale for the excitation is $\tau_e = 1 / \Gamma_e$. 

The ionization timescale of the high-Z ion $\tau_i$ in the blackbody photon bath is determined by the rate of ionization $\Gamma_i$, that is $\tau_i = 1 / \Gamma_i$. Considering the ionization to all energies above the threshold, $\Gamma_i$ can be written as~\citep{Kusenko_2012PhLB..707..255K}
\begin{equation}
\Gamma_{i} = \frac{64 \alpha \pi^3 k_{\mathrm{B}}^3}{3 c^2 h^3} a_Z^2 T^3 Y\left(\frac{\gamma}{\gamma_Z}\right)
\label{eq:Gamma_i}
\end{equation}
where
\begin{equation}
\begin{aligned}
& Y(r)=\int_{w=1}^{\infty} s(w) \chi\left(\frac{w}{r}\right) \frac{d w}{w^2}, \\
& s (w) = \frac{128}{w^3} \frac{1}{1 - \exp(-2 \pi / \sqrt{w - 1})} \\
& \times \exp (- \frac{2}{\sqrt{w - 1}} \arctan \frac{2 \sqrt{w - 1}}{2 - w}).
\end{aligned}
\end{equation}
$w = \Delta / \epsilon_z$ is the dimensionless parameter, and $\Delta$ is the ionization energy.

The cross-section of the recombination for the hydrogen-like ion to the energy state of $n$ is~\citep{Rybicki_1986rpa..book.....R}
\begin{equation}\label{sigma_rb}
    \sigma_{\rm rb} (n) = \frac{128 \pi^4 e^{10} Z^4 g_R (n)}{3 \sqrt{3} m_e c^3 h^4 \nu n^3 v^2},
\end{equation}
where $g_R (n) \simeq 1$ is the Gaunt factor for recombination, $h \nu$ is the energy of the radiation associated with the recombination, and $v$ is the velocity of the free electron. The timescale of the recombination at the lab frame then can be derived by using
\begin{equation}
    \tau_r = (n_e \sigma_{\rm rb} c)^{-1},
\end{equation}
where $n_e = L / \left[4 \pi (1 + \sigma) \gamma m_p R_{\rm emit}^2 c^3\right]$ is the number density of the electrons at the lab frame with $L$ being the power of the jet, and $\sigma \simeq 100$ is the magnetization factor before dissipation phase~\citep[e.g.,][]{Begue_2015ApJ...802..134B,Gao_2015ApJ...801..103G}. 

We should consider the propagation distance $d \simeq R_{\rm diss} - R_{\rm emit}$ for the high-Z ions. 
When the requirement (i.e., $d / c > \tau_e$) is satisfied and considering the de-excitation of the excited state (compared to the excitation) and the recombination of the ionized electron (compared to the ionization) are practically instantaneous~\citep{Kusenko_2012PhLB..707..255K}, the number of emitting photons with energy of $E_{\rm line}$ per ion per second in lab frame is $\Gamma_e$.
Considering the Doppler boosting caused by the relativistic motion of the shell, the observed isotropic luminosity of the emission line then can be derived
\begin{equation}
    L_{\rm line} = \frac{8 \gamma^2}{\theta_j^2} E_{\rm line} N_i \Gamma_e,
\end{equation}
where $N_i = M_i / m_i$ is the number of ions generating the observed emission line at a certain time, $m_i$ is the ion mass, and $\theta_j$ is the half opening-angle of the GRB jet.

On the other hand, for a hydrogen-like high-Z ion propagating within a distance $d$, the total number of photons that the ion emits is
\begin{equation}
    N_\gamma = \Gamma_e \frac{d}{c}.
\label{eq:N_tot}
\end{equation}
We then can derived the observed isotropic total energy $\mathcal{E}_{\rm line} = L_{\rm line} \Delta t_{\rm line}$ of the emission line by using
\begin{equation}
    \mathcal{E}_{\rm line} = \frac{4}{\theta_j^2} E_{\rm line} N_{i, \rm tot} N_\gamma,
\label{eq:N_toti}
\end{equation}
where $\Delta t_{\rm line}$ is the duration of the emission line at observer frame, $N_{i, \rm tot} = M_{i, \rm tot} / m_i$ is the total number of the high-Z ion generating the emission lines throughout $\Delta t_{\rm line}$. We can then get the total mass of the heavy nuclei carried in the jet from the observed isotropic total energy $\mathcal{E}_{\rm line}$ of the emission line, which is
\begin{equation}
    M_{\rm tot} = M_{i, \rm tot} / \zeta_i, % \zeta_\beta
\label{eq:M_tot}
\end{equation}
where 
$\zeta_{i} \sim 0.1$ means the mass fraction that can produce emission lines to the mass of all the nuclei that can be entrained in the GRB jet.

\section{MeV Emission Line in GRB~221009A}
\label{sect:221009a}

\begin{figure}[hbtp]
    \centering
    \subfigure[]{
    \includegraphics[width=0.45\textwidth]{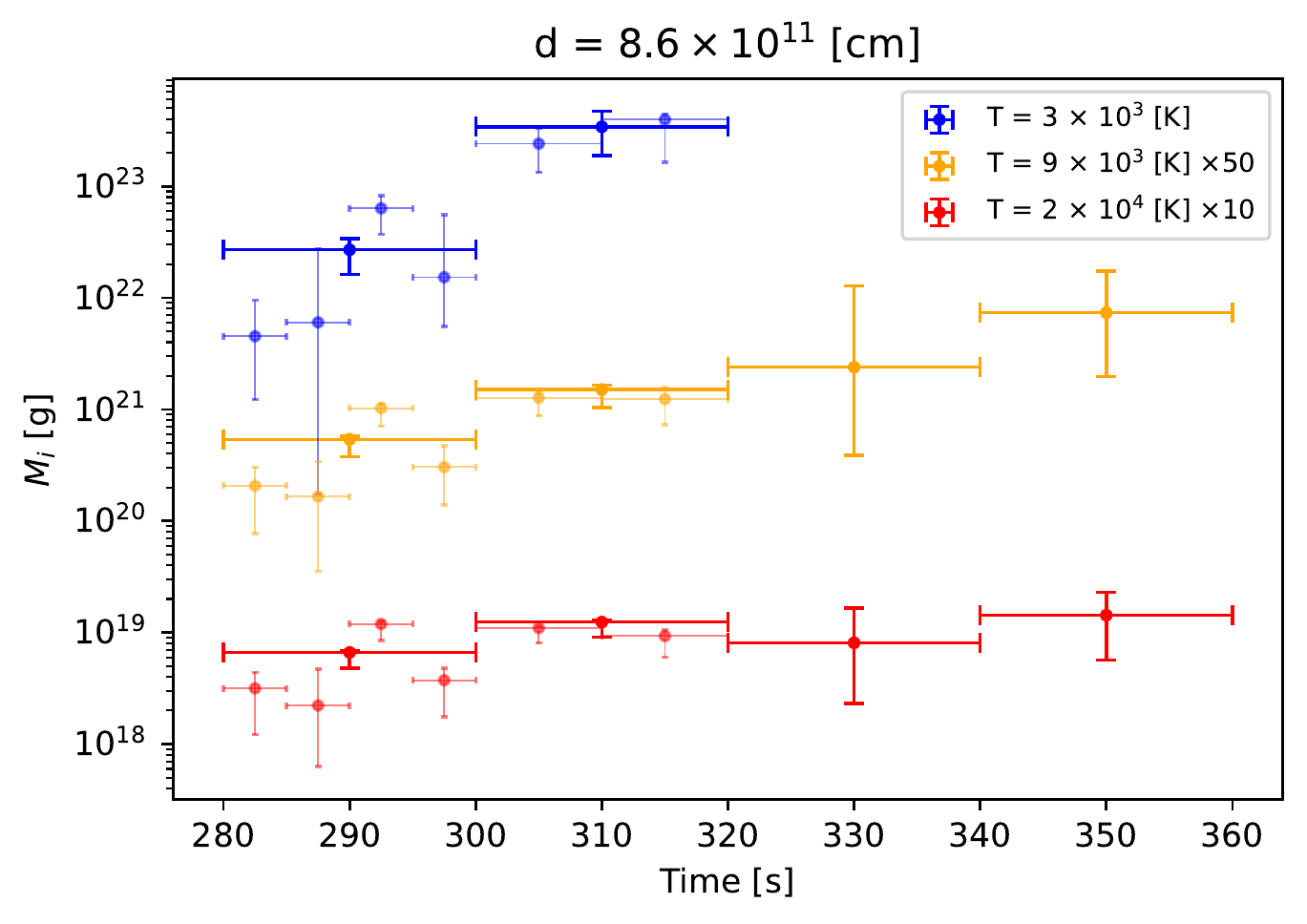}
    \label{fig:xi1}}
    \subfigure[]{
    \includegraphics[width=0.45\textwidth]{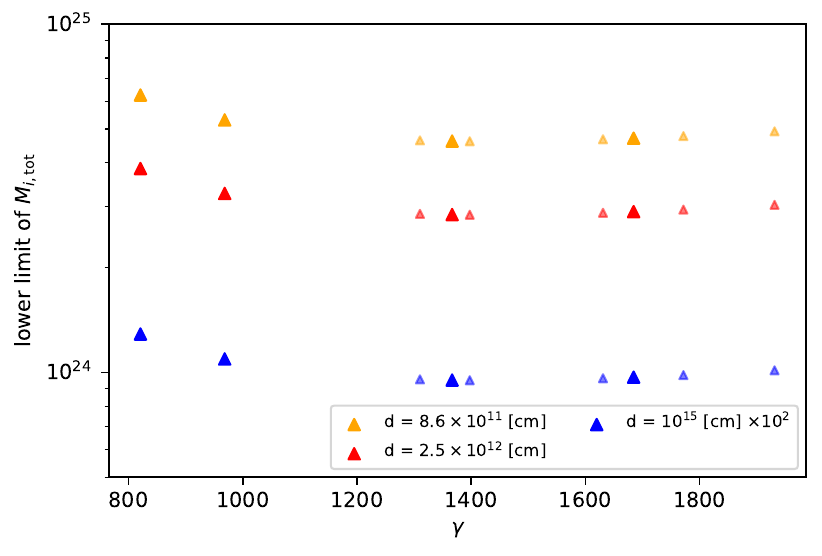}
    \label{fig:xi_low}}

    \caption{\textbf{Panel (a):} The value of $M_i$ contributing to the emission line at each time interval with $T = 3000 \rm~K$ (the blue point), $T = 9000 \rm~K$ (the orange point), and $T = 2 \times 10^4 {\rm~K}$ (the red point). The coefficient in the caption means the scaling of $M_i$.
    The other parameters we used are $d = 8.6 \times 10^{11} {\rm~cm}$ and $\theta_j = 0.02 {\rm~radian}$. 
    \textbf{Panel (b):} The lower limit of $M_{i, \rm tot}$ for $T = 10^4~$K with different $\gamma$ for $d = 8.6 \times 10^{11} {\rm~cm}$ (the orange triangle) corresponding to $\Delta t_{\rm gap} = 33~\rm s$, and $d = 2.5 \times 10^{12} {\rm~cm}$ (the red triangle) corresponding to $\Delta t_{\rm gap} = 97~\rm s$.
    We also show the results with $d = 10^{15} {\rm~cm}$ (the blue triangle) due to the uncertainty of the beginning time.} 
    \label{fig:xi}
\end{figure}

According to \cite{Edvige_2023arXiv230316223E}, the duration without the observation of emission lines is $\Delta t_{\rm gap} \simeq (33-97)$~s, where we take the the second time interval ($183~$s since GBM trigger) as the beginning time.
Note that the range of $\Delta t_{\rm gap}$ is due to the uncertainties at the time interval affected by saturation (called Bad Time Interval, BTI) which is $(219 - 277)$~s since GBM trigger. 
The narrow MeV emission lines which appear after the BTI lasted over 80~s with a roughly constant width $\sigma \sim 1$~MeV, a temporal evolving observed energy of $E_{\rm line} \simeq (12.56 - 6.12)$~MeV and a temporal evolving luminosity $L_{\rm line} \simeq (1.12 \times 10^{50} - 2.1 \times 10^{49}) {\rm~erg~s^{-1}}$~\citep{Edvige_2023arXiv230316223E}. 
Here we consider the MeV emission lines in GRB~221009A are generated by the excitation and de-excitation of the relativistic hydrogen-like copper. We take the copper for an example since large amounts of the nickel can be synthesized in the core-collapse process~\citep[e.g.,][]{Obergaulinger_2023arXiv230312458O} and some isotopes of nickel decay to copper with $\tau_{\beta,1/2} \lesssim (1 - 10^{3})$~s~\citep[e.g.,][]{Bazin_2017PhyOJ..10..121B,Audi_2003NuPhA.729....3A}.

Combined with Eq.~\ref{eq:E_line}, the Lorentz factor of the copper in the dominated jet shell is estimated to be $\gamma \approx 820 - 1700$. 
Although the required Lorentz factor is larger than the inferred Lorentz factor $\gamma \simeq 600$ for the narrow Poynting-dominated jet~\citep{Zhang_2023arXiv231113671Z}, it can be explained by the dissipation process that generates the prompt emission and afterglow emission, in which the kinetic energy of the jet transferred into the internal energy of the electrons, protons, and nuclei. Based on \cite{Ren_2023arXiv231015886R}, the dissipation efficiency may exceed 50\%.

The WR star is considered to be the progenitor for long GRBs~\citep[e.g.,][]{Zhang_2018pgrb.book.....Z}, with a typical temperature of $T_c \sim (3 \times 10^4 - 1.5 \times 10^5)$~K and a typical radius of $R_c \sim (10^{11} - 10^{12})$~cm~\citep[e.g.,][]{Beals_1933Obs....56..196B,Crowther_2007ARA&A..45..177C}. The reflection of the radiation from WR star can generate enough seed photons to excite electrons. 
Assuming the temperature of the isotropic seed photons from $3 \times 10^3$~K to $2 \times 10^4$~K, we can obtain the required $M_i$, as shown in Fig.~\ref{fig:xi1}. 
Note that the half-opening angle we used is $\theta_j \simeq 0.02~\rm radian$~\citep{Ren_2023arXiv231015886R}.
We can see that the temperature for the seed photons is more likely to be larger than $\sim 9000 {\rm~K}$ since the requirement $d / c > \tau_e$ can not be satisfied for $T = 3000$~K with a relatively low $\gamma$.  
Therefore, with a suitable distance $d_{\rm sc} \sim 10^{15}$~cm of scatterer and $R_{\rm emit} < d_{\rm sc}$, the reflection of the radiation from the WR star with $T_c \sim 10^5$~K and $R_c \sim 10^{12}$~cm can provide sufficient seed photons to generate MeV emission line.

We can estimate the total isotropic energy $\mathcal{E}_{\rm line}$ by integrating the luminosity of the emission line, which is $\simeq 4.3 \times 10^{51}$~erg. Note that if the emission line is present during the BTI time interval, the value of $\mathcal{E}_{\rm line}$ will be larger. 
Based on Eq.~\ref{eq:N_toti}, the constraints for $M_{i, \rm tot}$ with a fixed value $T$ then can be derived, as shown in Fig.~\ref{fig:xi_low}. 
The different color means different $d$, as indicated in the caption.
Note that the required $M_{i, \rm tot}$ is smaller for a larger $d$ due to a longer time for the copper to generate the emission line.
Combined with Eq.~\ref{eq:M_tot}, we can then derive the lower limit of the total mass of heavy nuclei entrained in the GRB jet, which is $M_{\rm tot} \sim (10^{23} - 10^{26}) {\rm~g}$, corresponding to a total number of $10^{45} - 10^{48}$.
The total mass of the GRB jet can be estimated to be $ M_{\rm jet} =  \frac{\theta_j^2 \mathcal{E}_{\rm iso}}{4 \Gamma_0 c^2} \approx 2.98 \times 10^{27}$~g, where $\mathcal{E}_{\rm iso} \simeq 1.5 \times 10^{55} {\rm~erg}$ and $\Gamma_0 \simeq 560$ are the intrinsic isotropic equivalent energy and the initial Lorentz factor of external shock for GRB~221009A, respectively~\citep[e.g.,][]{Frederiks_2023ApJ...949L...7F,Burns_2023ApJ...946L..31B,LHAASO_2023Sci...380.1390L}. 
The mass fraction of $M_{\rm tot}$ to $M_{\rm jet}$ is $\xi_{\rm tot} \sim 10^{-4} - 10^{-1}$.

\section{Summary and Discussion}
\label{sect:sum}
In this work, considering the excitation and de-excitation of the hydrogen-like high-Z 'cold' ions in the relativistic jet, we have proposed a model for generating MeV narrow emission lines in the observed spectra of prompt radiation. Based on our model and the emission line observations, we can constrain 
the total mass of the heavy nuclei carried in the jet ($M_{\rm tot}$), and the coasting phase Lorentz factor ($\gamma$). 

We applied our model to the MeV emission lines in GRB~221009A. Despite the lack of observations during the BTI phase, we can give some constraints on the properties of GRB:
(1) The coasting phase Lorentz factor of the jet is estimated to be $\gamma \approx  820 - 1700$, and the temperature $T$ of blackbody seed photons is $\gtrsim 9000$~K; 
(2) The lower limit of the total mass of heavy nuclei entrained in GRB jet, that is $M_{\rm tot} \sim (10^{23} - 10^{26}) {\rm~g}$, corresponding to a mass fraction of $\xi_{\rm tot} \sim 10^{-4} - 10^{-1}$.
Our model can well explain the MeV emission line of GRB~221009A with reasonable parameters.

However, there is only one event (i.e., GRB~221009A) that has observed the MeV emission line so far. It might be due to the strict requirements for its generation: (1) The parameter space that can generate emission lines is relatively small; (2) The relatively high temperature of the progenitor $T \sim 10^5$~K and the relatively close distance of the scattering cold cloud $d_{\rm sc} \sim 10^{15}$~cm are required to provide enough seed photons; (3) The jet is better to be Poynting-flux-dominated so that the jet can carry enough heavy nuclei in the jet base. Similar narrow emission lines are possible only for GRBs satisfying the above requirements.

Finally, we can give a prediction for future observations. 
A bump at $\rm GeV-TeV$ band might be produced by the Inverse Compton (IC) process between the shocked electrons and the photons from emission lines since the emission line photons need to break through the shocked region at $\gtrsim R_{\rm diss}$. 
Also, the existence of significant heavy nuclei in low-luminosity GRBs is quite possible~\citep[e.g.,][]{Horiuchi_2012ApJ...753...69H, Zhang_2018PhRvD..97h3010Z}, which might be related to a sub-MeV emission line due to a relatively smaller $\gamma < 100$~\citep[e.g.,][]{Zhang_2012ApJ...756..190Z,Zhang_2021ApJ...920...55Z}.

\section*{Acknowledgments}
We thank helpful discussions with Bing Zhang, Zhi-Ping Jin, Kohta Murase, Lu-Yao Jiang, Yi-Ping Li, Hao Zhou, Xiang-Yu Wang, Felix Aharonian, Abe Falcone, and Tie-Kuang Dong.
D.M.W is supported by the National Natural Science Foundation of China (No. 11933010, 11921003, 12073080, 12233011) and the Strategic Priority Research Program of the Chinese Academy of Sciences (grant No. XDB0550400). Z.G.D is supported by the National SKA Program of China (grant No. 2020SKA0120300) and National Natural Science Foundation of China (grant No. 12393812). Y.P.Y is supported by the National Natural Science Foundation of China grant No.12003028 and the National SKA Program of China (2022SKA0130100). 
H.N.He is supported by Project for Young Scientists in
Basic Research of Chinese Academy of Sciences (No. YSBR-
061), and by NSFC under the grants No. 12173091, and No. 12333006.

\textbf{A note added in the proof:} Recently, \cite{Zhang_2024SCPMA..6789511Z} found the MeV emission line in GRB~221009A, which also can be explained by our model.

\appendix

\section{A More Detailed Description of Model}

%\subsection{Where Could a Jet Carry Heavy Nuclei?}
\subsection{Origins of Heavy Nuclei in a Jet}
\label{sect:where}

We consider two possible mechanisms to generate a jet carrying heavy nuclei.
The first mechanism is a jet could entrain heavy nuclei from the stellar Fe core.
At the last time of a massive star, the nuclear fusion process in the stellar center will produce unstable nuclei heavier than irons, and eventually lead to the collapse process, even a jet can be produced.
The initial fireball of jet with $r \sim (10^6 - 10^7) {\rm~cm}$ and isotropic energy $E_{\rm iso} \sim 10^{52}$~erg is too hot ($kT = (1 - 10)$~MeV), which is comparable to the nuclear binding energy ($\sim 10$~MeV), so that heavier nuclei will be broken down due to the photodisintegrated. 
However, the fireball temperature will decrease significantly as the jet expands, with $T \propto r^{-1}$. When the fireball reaches $\sim 10^9 {\rm ~cm}$ which is the radius of the stellar Fe core, the temperature of the fireball is no longer able to disintegrate the heavy nuclei. Thus the jet could entrain some of the heavy nuclei at this time.
However, with the accretion of the black hole, the material in the stellar core is continuously consumed, and the original heavy nuclei may soon be emptied. 
Also, for GRB~221009A, the narrow jet is Poynting-flux-dominated~\citep{Zhang_2023arXiv231114180Z}, and it is difficult for the heavy nuclei in the stellar envelope to enter this narrow jet since the strong magnetic field will block charges away~\citep{Lei_2013ApJ...765..125L}. Although it might be possible for the heavy nuclei to be entrained by a matter-dominated structured jet wing that is around the narrow jet, the Lorentz factor of the jet wing is too small to generate a MeV emission line based on our consideration.

The second mechanism is that a jet could carry heavy nuclei from the accretion disk wind of a hyperaccretion BH, i.e., a neutrino-dominated accretion flow (NDAF).
For such a disk, the temperature is high enough to ignite nuclear fusion and drive strong outflows.
The abundance of free protons and neutrons in the disk outflows will participate in nucleosynthesis or even induce core-collapse supernova explosions.
When mass accretion rates above a critical threshold value ($(10^{-3}-10^{-1}) ~\rm M_{\odot}~ s^{-1}$), 
which depends on the effective viscosity and BH mass~\citep{Chen_2007ApJ...657..383C,Metzger_2008MNRAS.390..781M}, the inner regions of the disk are ``self-neutralize"
via electron captures on protons~\citep{Beloborodov_2003ApJ...588..931B}, 
thus maintaining a low electron fraction $Y_e \sim 0.1$ in a regulated process~\citep{Siegel_2017PhRvL.119w1102S}. 
As a result, the collapsar disk
outflows, which feed on this neutron-rich reservoir, can
themselves possess a sufficiently high neutron concentration to
enable an r-process, during much of the epoch in which the GRB jet is being powered.
The disk outflows are tightly bound to a narrow region near the disk by stellar material (and will be expanding during the material being consumed with reduced density), which makes the jet potentially quite effective for the entrainment process of the heavy nuclei. 
According to \cite{Horiuchi_2012ApJ...753...69H} and \cite{Zhang_2018PhRvD..97h3010Z}, the heavy nuclei can survive at the location of the jet base $r_0 \sim 10^{9}$~cm which is similar to the critical radius of disk formation.
In this mechanism, the heavy nuclei from the disk can be entrained by a Poynting-flux-dominated jet from the jet base, which can avoid the issue of adding baryon loading from surroundings mentioned in \cite{Lei_2013ApJ...765..125L}. Also, based on the model including an intrinsically episodic jet that is launched from the disk through a magnetic process~\citep{Yuan_2012ApJ...757...56Y}, baryons from the disk can be directly entrained in the magnetic bubble, leading to the baryon-rich jet.

\subsection{Excitation Rate and Ionization Rate for a Given Photon Spectrum}

Here we rewrite the calculations on \cite{Kusenko_2012PhLB..707..255K} to calculate the excitation rate and ionization rate of any isotropic photon field, not limited to blackbody radiation. In the transition of the electron from the ground $1S$ to a $nP$ state, the excitation rate in an isotropic photon bath is
\begin{equation}
\begin{aligned}
    \Gamma_{e, n} & = \frac{16 \pi^3 \alpha \gamma}{3 c^2} \left( \sum_{\mathrm{pol}} |\langle nP | \vec{r} | 1S \rangle|^2 \right) \frac{\Delta_n^2}{h ^2\gamma^2} \int^\infty_{\Delta_n / (2 \gamma h)} n_{\nu} (\nu) d\nu \\
    & = \frac{2 \alpha \gamma}{3 c^2} \left( \sum_{\mathrm{pol}} |\langle nP | \vec{r} | 1S \rangle|^2 \right) \chi(\gamma, Z, \Delta_n),
\end{aligned}
\end{equation}
where
\begin{equation}
    \chi(\gamma, Z, \Delta_n) = 8 \pi^3 \frac{\Delta_n^2}{h ^2\gamma^2} \int^\infty_{\Delta_n / (2 \gamma h)} n_{\nu} (\nu) d\nu.
\end{equation}
The lower limit of the integral indicates the minimum frequency of the photons which can excite the ground electron to the n-level electron in a head-on collision. We can get the average number of photons $n_\nu (\nu)$ of frequency $\nu$ by using
\begin{equation}
    u_\nu (\nu) = \Delta \Omega h \nu n_\nu \rho_s,
\end{equation}
where $\rho_s = 2 \nu^2 / {c^3}$ is the density of states for an isotropic photon field, and $n_\nu (\nu)$ is the specific energy density.
For a simple hydrogen-like ion, the dipole matrix elements can be 
\begin{equation}
    \left( \sum_{\mathrm{pol}} |\langle nP | \vec{r} | 1S \rangle|^2 \right) = \frac{256 n^7}{(n^2 - 1)^5} \left(\frac{n-1}{n+1}\right)^{2n} \alpha_z^2 \ \ \ (n \geq 2).
\end{equation}
The total excitation rate should be
\begin{equation}
\label{eq:appe_gamma_e}
    \Gamma_e = \frac{2 \alpha}{3 c^2} \alpha_z^2 X(\gamma, Z),
\end{equation}
where
\begin{equation}
\begin{aligned}
    X(\gamma, Z) = & 256 \sum^\infty_{n=2} \frac{n^7}{(n^2 - 1)^5} \left( \sum_{\mathrm{pol}} |\langle nP | \vec{r} | 1S \rangle|^2 \right) \\
    & \times \chi(\gamma, Z, \Delta_n).
\end{aligned}
\end{equation}

Similar to Eq.~\ref{eq:appe_gamma_e}, the ionization rate $\Gamma_i$ in the isotropic photon bath can be written as
\begin{equation}
    \Gamma_i = \frac{2 \alpha \gamma}{3 c^2} \alpha_z^2 Y(\gamma, Z),
\end{equation}
where
\begin{equation}
\begin{aligned}
    & Y(\gamma, Z) = \int_{w = 1}^{\infty} s(w) \chi(\gamma, Z, \Delta), \\
    & \Delta = w \epsilon_z, \\
    & s (w) = \frac{128}{w^3} \frac{1}{1 - \exp(-2 \pi / \sqrt{w - 1})} \\
    & \times \exp (- \frac{2}{\sqrt{w - 1}} \arctan \frac{2 \sqrt{w - 1}}{2 - w}).
\end{aligned}
\end{equation}

\begin{figure}
	\centering
	\subfigure[]{
		\includegraphics[width=0.45\textwidth]{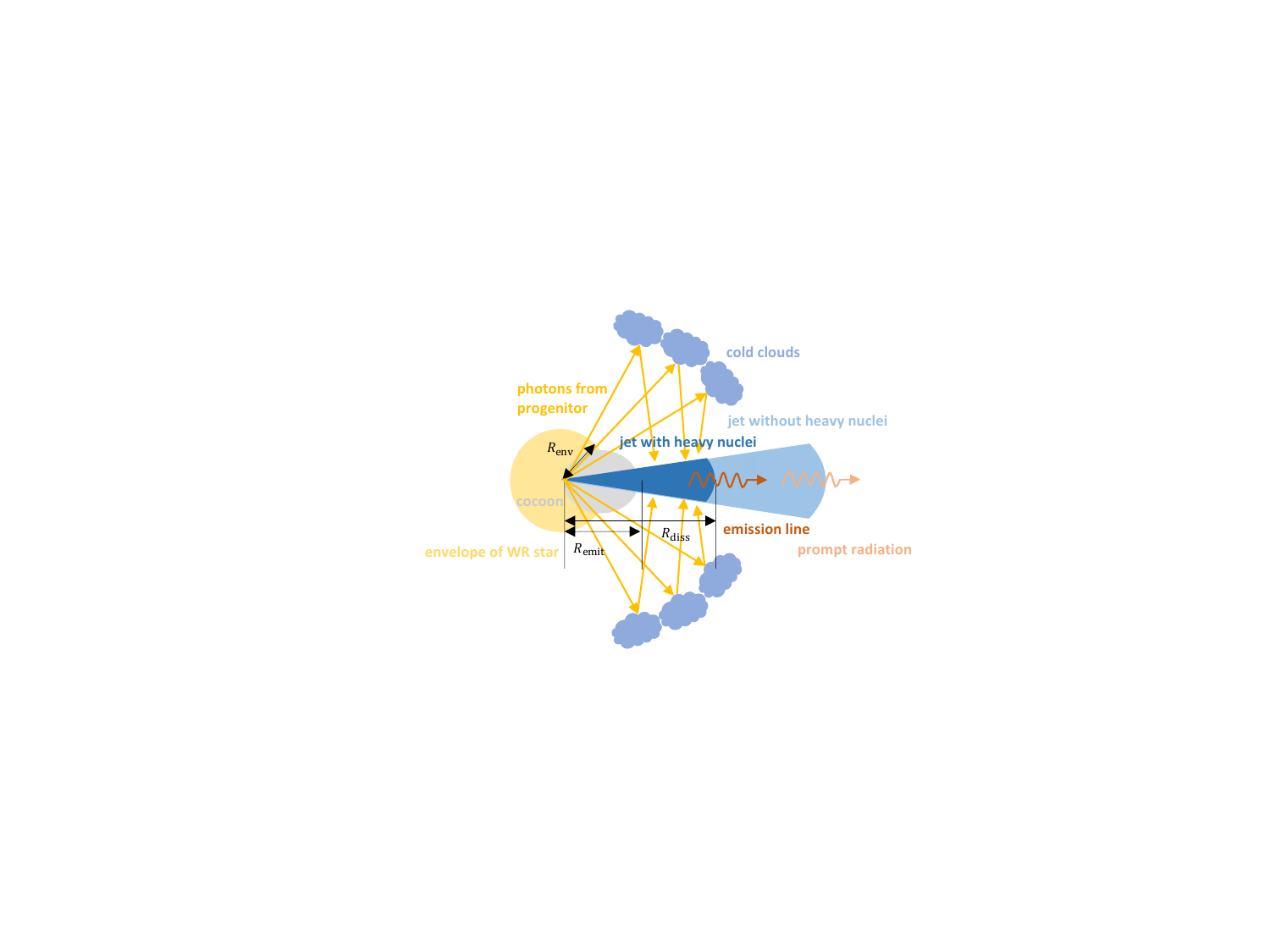}
		\label{fig:model_scatter}}
	\subfigure[]{
		\includegraphics[width=0.45\textwidth]{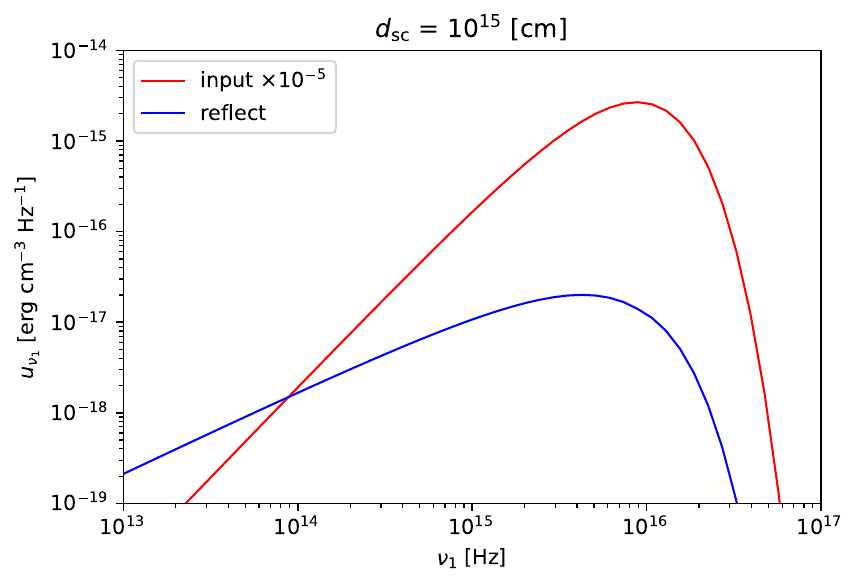}
		\label{fig:reflect}}
	
	\caption{\textbf{Panel (a):} A schematic model for scattering the photons from progenitor. The region that generates the emission line can have the almost isotropic seed photons from the reflection of the cold cloud shell since only within a small solid angle $\sim \theta^2 / 2$ the clouds might be destroyed by the jet. 
		\textbf{Panel (b):} The specific energy density for the radiation of the WR star (red line) and Compton reflection (blue line) at $d_{\rm sc} = 10^{15}$~cm with $\Delta \Omega = 1$~sr. The temperature of incident photons is $T = 1.5 \times 10^4$~K.} 
\end{figure}

\subsection{Possible Origins of Seed Photons}
\label{sect:photon}

First, we consider the influence of the photons in the jet. We use the photons of the internal shock to do the estimations since there are no more significant photons we can observe in the spectra of GRB~221009A. 
We can derive the luminosity of these photons $\simeq 6.57 \times 10^{50} {\rm~erg~s^{-1}}$ based on the spectra near the emission line (with a flux of $\sim 10^{-5} {\rm~erg~s^{-1}~cm^{-2}}$)~\citep{Edvige_2023arXiv230316223E}, and its luminosity distance $d_L \approx 2.1 \times 10^{27}$~cm. 
Combined with the location of the jet as $(10^{13} - 10^{15})$~cm, we can get the flux of these photons $(5.23 \times 10^{23} - 5.23 \times 10^{19}){\rm~erg~s^{-1}~cm^{-2}}$, corresponding to the effective temperature of $(9.8 \times 10^6 - 9.8 \times 10^5)$~K in the copper comoving frame if we assuming blackbody radiation. 
We have to transfer this temperature to the lab frame, and then do the calculations based on our model. 
We find that for $d \gtrsim 10^{14} {\rm~cm}$, the emitting photons per coppers $N_\gamma \ll 1$, indicating that the pre-existed photons are difficult to generate MeV emission lines and to ionize the electron of coppers. 
However, $N_\gamma \sim 10$ for $d = 10^{13}$~cm indicates that if the emission line occurs near the center, the photons in the jet can emit the MeV emission lines within the ionization time scale of $\tau_i \sim 10^4 - 10^5$~s. 
Note that the distance for generating the emission lines should be larger than $10^{12}$~cm, since $\tau_i \sim 10^{-3}$~s is too short, resulting in the lack of hydrogen-like coppers in the jet. 
According above calculations, we can see that the pre-existence photons can be another possible origin for the seed photons to generate an emission line which depends on the location of the emitting region. 
However, we do not find the corresponding blackbody radiation in the keV band for GRB~221009A, indicating the less possibility for the existence of photons in the jet at a radius of $R_{\rm emit} - R_{\rm diss}$. 

Another possible origin of seed photons is the reflection of the radiation from the GRB's progenitor.
The WR star is considered to be the progenitor for long GRBs~\citep[e.g.,][]{Zhang_2018pgrb.book.....Z}, with a typical temperature of $T_c \sim (3 \times 10^4 - 1.5 \times 10^5)$~K and a typical radius of $R_c \sim (10^{11} - 10^{12})$~cm~\citep[e.g.][]{Beals_1933Obs....56..196B,Crowther_2007ARA&A..45..177C}. 
Therefore, it can contribute to the emission line, and we consider the possibility of the seed photons generated by the reflection of the radiation of the WR star.
We assume the scatterer might be the quasi-spherical cold shell which is ejected by the explosion of the progenitor, and is located at a specific distance $d_{\rm sc} \sim (10^{13} - 10^{15})$~cm from the progenitor, as shown in Fig.~\ref{fig:model_scatter}. The cold cloud shell might be destroyed by the GRB jet only within a solid angle of $\theta^2 / 2$ considering the opening angle of the jet and the beaming effect. In our model, $d_{\rm sc}$ is larger than $R_{\rm emit}$, and the photons from the WR star are reflected by the cold cloud.
We can approximate our model as sphere-shell luminescence, and thus use the following equations to get the specific radiation energy density at $R_{\rm emit}$~\citep{Ghisellini_2013LNP...873.....G}
\begin{equation}
\begin{aligned}
        u_\nu & = \frac{\alpha_{\rm ref} L_{\nu, c} \Delta \Omega }{4 \pi d_{\rm sc}^2 c} \\
        & = \frac{2 \pi h \alpha_{\rm ref} R_c^2 \Delta \Omega }{d_{\rm sc}^2 c^3} \frac{\nu^3}{\exp{\left(\frac{h \nu}{k_B T_c}\right)} - 1},
\end{aligned}
\end{equation}
where $\alpha_{\rm ref} \sim 0.1$ is the albedo for cold clouds~\citep[e.g.,][]{Fritz_1949JAtS....6..277F}, $L_{\nu, c}$ is the specific luminosity of the blackbody radiation from WR star, $\Delta \Omega = A_{\rm sc} / d_{\rm sc}^2 \sim 1$~sr is the solid angle for the cold cloud shell with $A_{\rm sc}$ being the effective area of it.
Note that the reflection will cause changes in the spectrum. For simplicity, we assume the optical depth of the cold clouds is larger than one, and the reflection is mainly through the Compton scattering process by cold electrons in a spherical medium~\citep[e.g.,][]{Lightman_1981ApJ...248..738L,White_1988ApJ...331..939W}. The final reflection energy density $u_{\nu_1} (\nu_1)$ at frequency $\nu_1$ at $R_{\rm emit}$ is
\begin{equation}
    u_{\nu_1} (\nu_1) = \int G(\nu, \nu_1) u_{\nu} (\nu) d \nu,  
\end{equation}
where $G(\nu, \nu_1)$ is the fitted Green's function in the nonrelativistic limit ($x = h \nu / \left(m_e c^2\right) \ll 1$)
\begin{equation}
\begin{aligned}
        G(\nu, \nu_1) \approx & \alpha_{n} \frac{h}{m_e c^2} x^{-2} \\
        & \times 
        \begin{cases}
          0.10 & 1/x_1 - 1/x < 2 \\
          0.56 \left(1/x_1 - 1/x\right)^{-3/2} & 1/x_1 - 1/x \geq 2.
        \end{cases}
\end{aligned}
\end{equation}
Note that $\alpha_n$ is the normalization factor to ensure the photon number is conserved.
We use the blackbody radiation from WR star with $T_c = 1.5 \times 10^5$~K to calculate the reflection specific energy density $u_\nu$ at $R_{\rm emit}$ with a scatterer at $d_{\rm sc} = 10^{15}$~cm, as shown in the blue line of Fig.~\ref{fig:reflect}. The red line shows the value of $u_\nu$ from the WR star with $T_c = 1.5 \times 10^5$~K and $R_c = 10^{12}$~cm. Note that in our approximation, the photon density is the same in the region within the scattering screen. Therefore, the requirement for $R_{\rm emit}$ is only $R_{\rm emit} < d_{\rm sc}$.
However, one needs to rewrite Eqs.~\ref{eq:Gamma_e} and \ref{eq:Gamma_i} to calculate the luminosity of the emission line since the seed photons no longer follow a blackbody radiation distribution. We need to replace $\left(2 \pi k T / h\right)^3 \chi(z)$ in Eq.~\ref{eq:Gamma_e} and Eq.~\ref{eq:Gamma_i} with the following form
\begin{equation}
2 \pi \left(\frac{2 \pi \Delta_n}{h \gamma}\right)^2 \int_{\frac{\Delta_n}{2 \gamma h}}^{\infty} \frac{c^3 u_\nu (\nu)}{2 h \nu^3} d \nu,
\end{equation}
where $\Delta_n = \left(1 - n^{-2}\right) \epsilon_z$ is the excitation energy of the n-level electron transition back to the ground state, and $n_\nu (\nu)$ is the average number of photons of frequency $\nu$. Combined with the specific energy density $u_{\nu}$ of reflected photons, one can calculate the luminosity of the emission line with the modified Eqs.~\ref{eq:Gamma_e} and \ref{eq:Gamma_i}.
We find that, with a suitable $d_{\rm sc} \sim 10^{15}$~cm, the reflection of the radiation from the WR star with $T_c \sim 10^5$~K and $R_c \sim 10^{12}$~cm can provide sufficient seed photons to generate MeV emission line.

\section{Parameter Space for Emission Line}
\label{sect:predict}

\begin{figure*}
    \centering
    \subfigure[]{
    \includegraphics[width=0.35\textwidth]{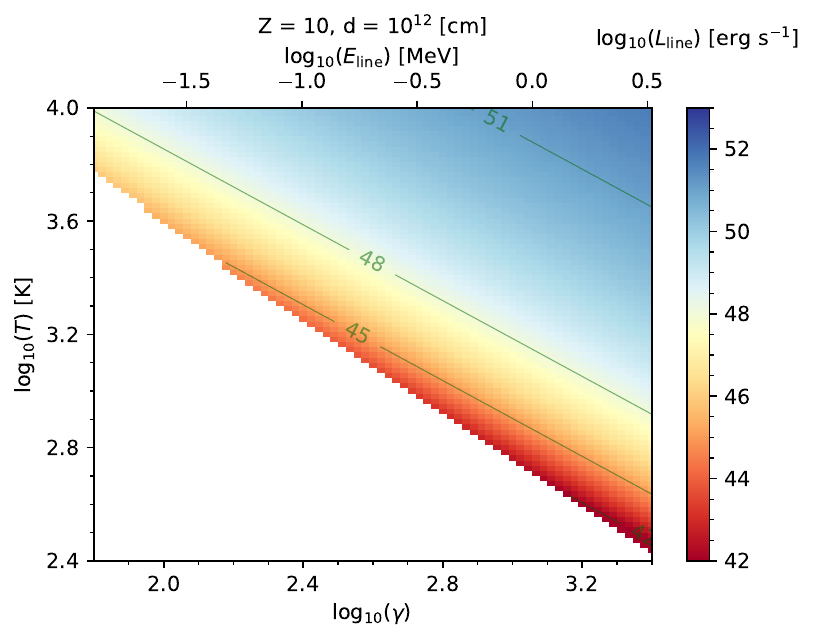}
    \label{fig:a}}
    \subfigure[]{
    \includegraphics[width=0.35\textwidth]{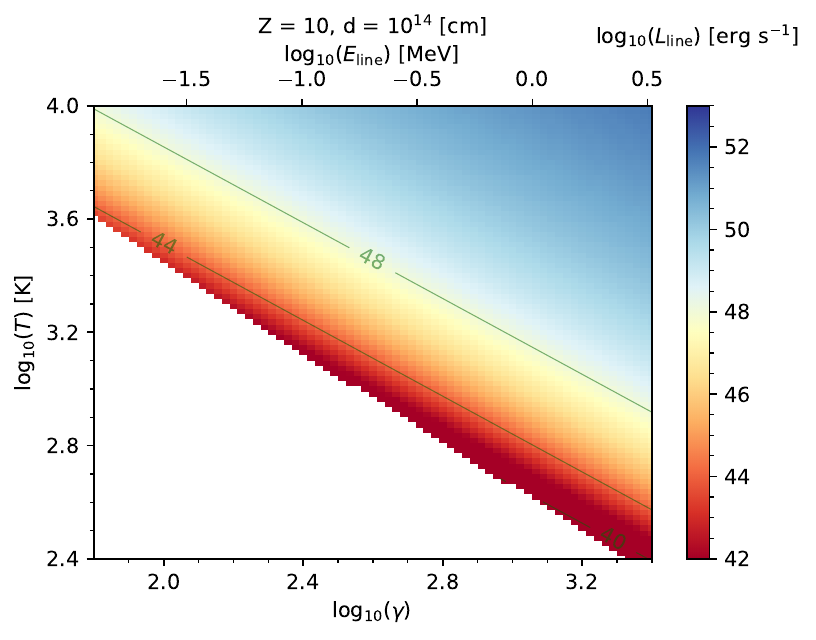}
    \label{fig:b}}

    \subfigure[]{
    \includegraphics[width=0.35\textwidth]{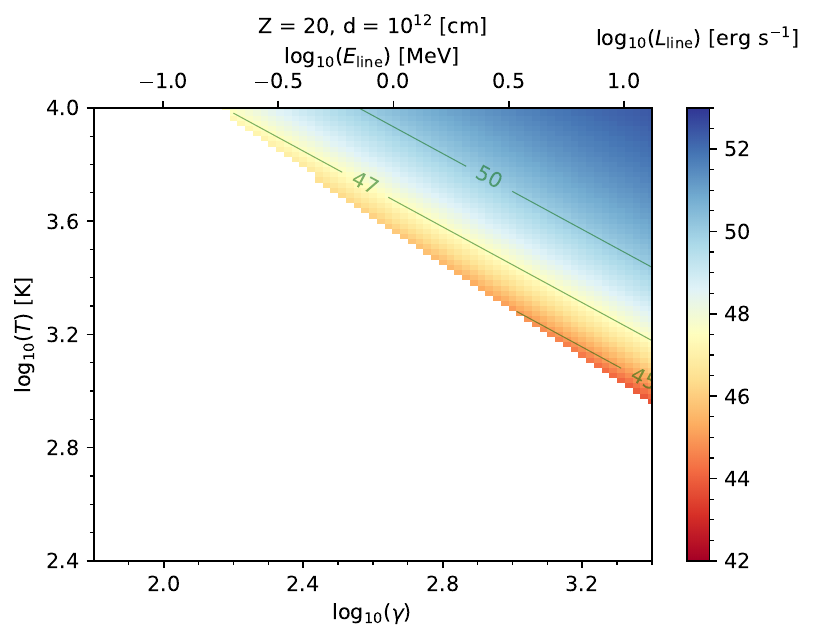}
    \label{fig:c}}
    \subfigure[]{
    \includegraphics[width=0.35\textwidth]{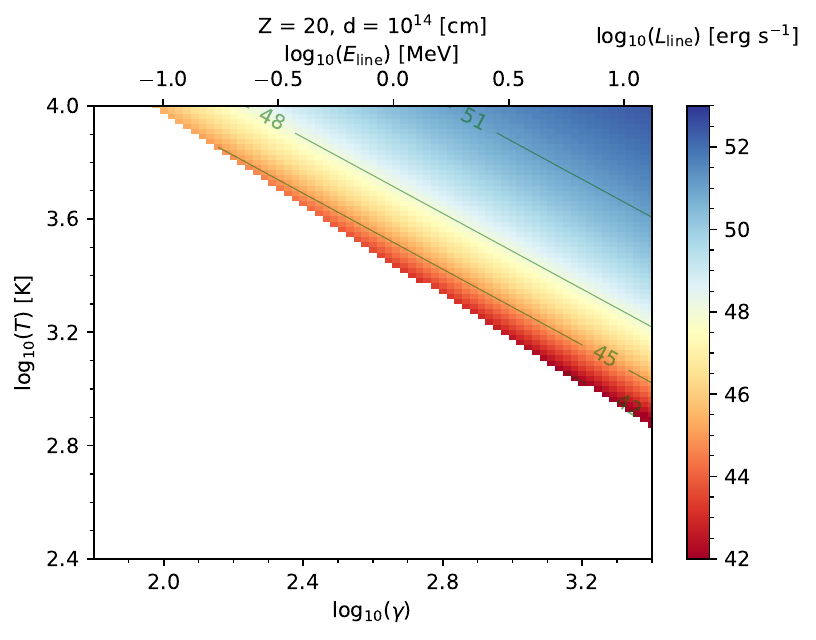}
    \label{fig:d}}

    \subfigure[]{
    \includegraphics[width=0.35\textwidth]{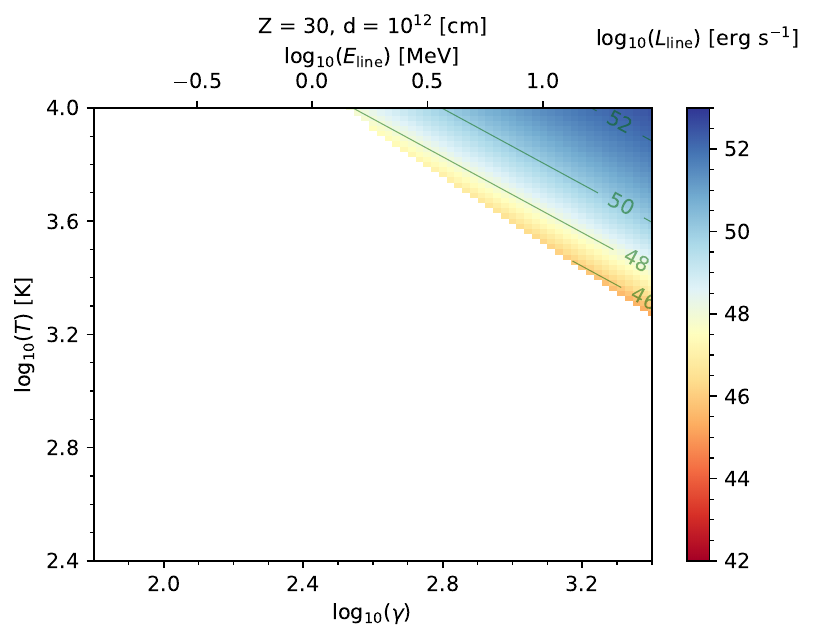}
    \label{fig:e}}
    \subfigure[]{
    \includegraphics[width=0.35\textwidth]{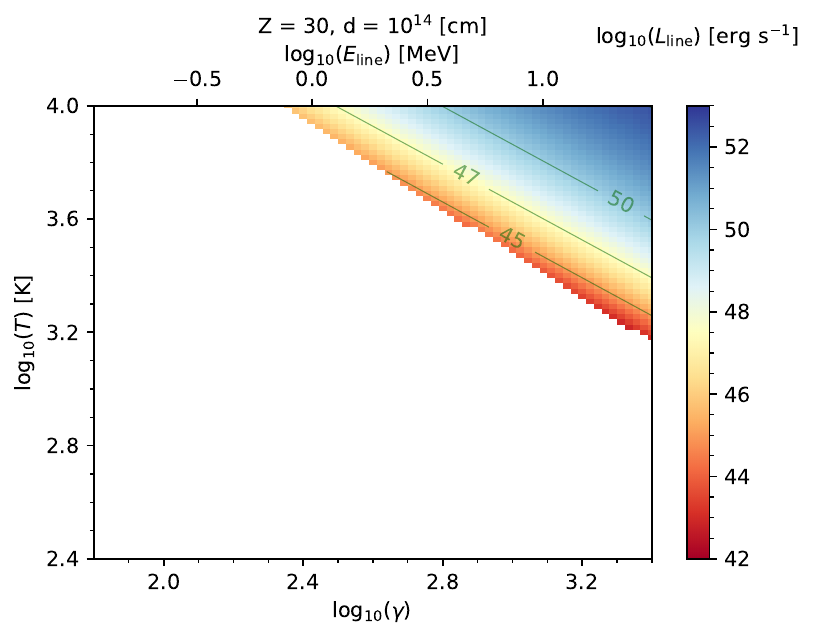}
    \label{fig:f}}

    \subfigure[]{
    \includegraphics[width=0.35\textwidth]{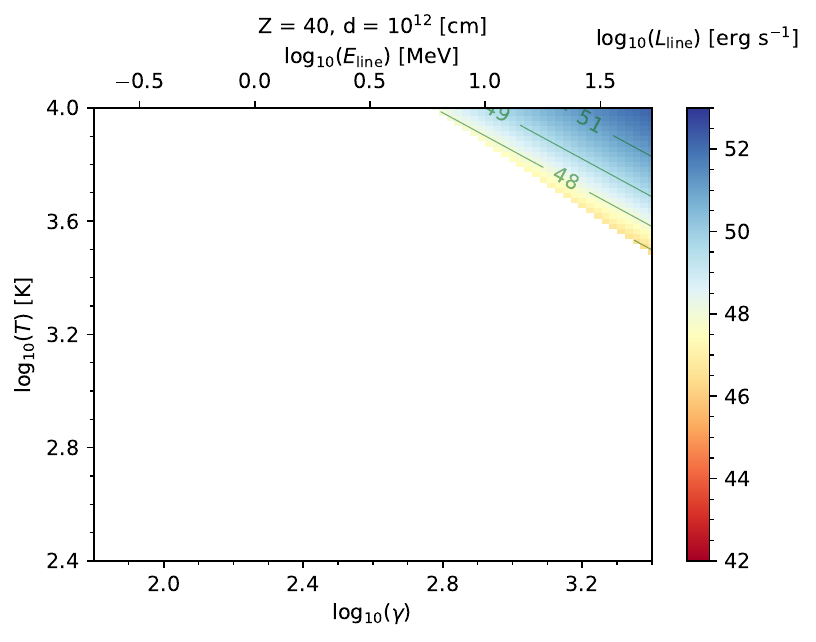}
    \label{fig:g}}
    \subfigure[]{
    \includegraphics[width=0.35\textwidth]{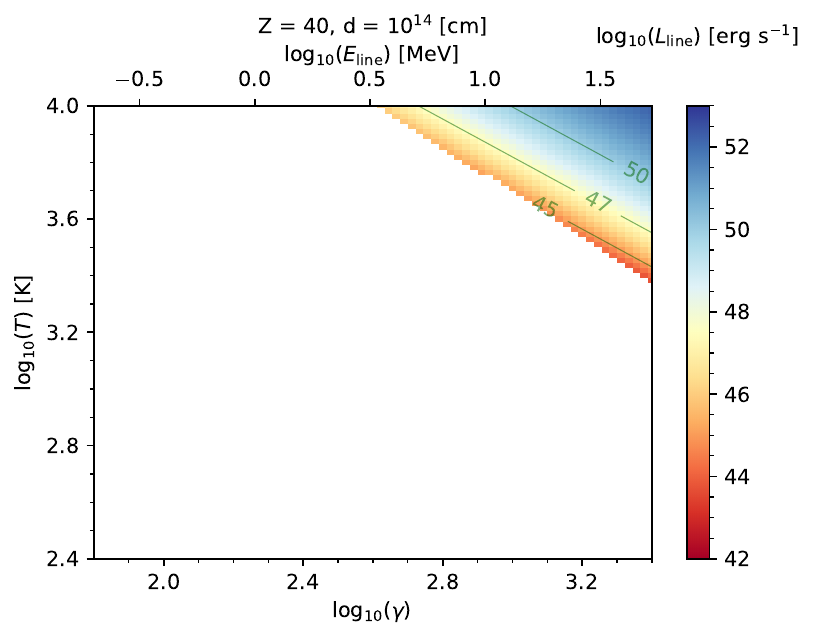}
    \label{fig:h}}

    \caption{The parameter space in the $\gamma$ -- $T$ plane with the color indicating the isotropic luminosity with $\theta_j = 0.01$~rad and $N_i = 10^{42}$~counts. Blank areas indicate that this part of the parameter space cannot generate emission lines. We use $d = 10^{12}$~cm ($10^{14}$~cm) in the left (right) panel, which mainly affects the size of the parameter space but not the luminosity. We use $Z = 10$, $20$, $30$, and $40$ in Figs.~\ref{fig:a}-\ref{fig:f}, which affects both the parameter space and the luminosity.} 
    \label{fig:para_space}
\end{figure*}

In Fig~\ref{fig:para_space}, we scan the parameter space to find the restrictions on the parameters of our model. 
We fixed $\theta_j = 0.01$~rad and $N_i = 10^{42}$~counts to calculate $L_{\rm line}$. Note that the values of $N_i$ and $\theta_j$ only change the luminosity of the emission line ($L_{\rm line} \propto N_i \theta_j^{-2}$) but not the contour shape of allowed parameter space. 
In the left (right) panel, we use $d = 10^{12}$~cm ($d = 10^{14}$~cm). The allowed parameter space with $d = 10^{12}$~cm is smaller than that with $d = 10^{14}$~cm, but their luminosity is the same. In Figs.~\ref{fig:a}-\ref{fig:h}, we use $Z = 10$, $20$, $30$, and $40$, respectively. The value of $Z = 30$ and $40$ we choose indicates the peaks of nucleosynthesis for heavy nuclei~\citep[e.g.,][]{Wanajo_2014ApJ...789L..39W,Obergaulinger_2023arXiv230312458O}. We use $Z = 10$ and $20$ to estimate the possible product of the accretion for the helium star. The allowed parameter space and the luminosity with a smaller $Z$ are larger. Note that the energy of emitted photons from heavy nuclei is $\propto \gamma Z^2$ so that the emission line may exist in different energy bands for different values of $Z$ and $\gamma$.

Based on the calculations on \cite{Horiuchi_2012ApJ...753...69H} and \cite{Zhang_2018PhRvD..97h3010Z}, the existence of significant heavy nuclei in low-luminosity GRBs is quite possible, which might be related to a sub-MeV emission line due to a relatively smaller $\gamma < 100$~\citep[e.g.,][]{Zhang_2012ApJ...756..190Z,Zhang_2021ApJ...920...55Z}. However, in our calculations, we find that such sub-MeV emission line is difficult to be detected due to its relatively low $\gamma$, and only with $Z \lesssim 20$, $d \gtrsim 10^{14}$~cm, and $T \gtrsim 10^4$~K, the GRB jet with $\gamma \lesssim 100$ can generate emission lines at the sub-MeV band.


\bibliography{sample631}{}
\bibliographystyle{aasjournal}


\end{CJK*}
\end{document}